\documentclass[twocolumn,superscriptaddress,floatfix,preprintnumbers,prd,nofootinbib,aps,10pt]{revtex4-2}

\usepackage[colorlinks=true,breaklinks=true]{hyperref}
\hypersetup{allcolors=[rgb]{0.0 0.0 0.7},linkcolor=[rgb]{0.75 0.05 0.05}}
\usepackage[dvipsnames]{xcolor}
\usepackage{mathtools}
\usepackage{amsmath}
\usepackage{amsfonts}
\usepackage{amssymb}
\usepackage{bbold}
\usepackage{bm}
\usepackage{orcidlink}
\usepackage{microtype}

\newcommand{\bQ}{{\bf Q}}

\newcommand{\bA}{{\bf A}}
\newcommand{\bH}{{\bf H}}
\newcommand{\bB}{{\bf B}}
\newcommand{\bP}{{\bf P}}
\newcommand{\bR}{{\bf R}}
\newcommand{\bS}{{\bf S}}

\newcommand{\bL}{{\bf L}}
\newcommand{\bJ}{{\bf J}}
\newcommand{\bK}{{\bf K}}
\newcommand{\bG}{{\bf G}}

\newcommand{\half}{{\textstyle\frac{1}{2}}}

\newcommand{\vc}{v_{\rm c}}

\newcommand{\GF}{G_{\rm F}}
\long\def\exclude#1{}

\begin{document}

\title{The ubiquitous flavor pendulum}

\author{Damiano F.\ G.\ Fiorillo \orcidlink{0000-0003-4927-9850}}
\affiliation{Istituto Nazionale di Fisica Nucleare (INFN), Sezione di Napoli, Complesso Universitario di Monte Sant'Angelo, Via Cintia, 80126 Napoli, Italy}
\affiliation{Deutsches Elektronen-Synchrotron DESY,
Platanenallee 6, 15738 Zeuthen, Germany}

\author{Georg G.\ Raffelt
\orcidlink{0000-0002-0199-9560}}
\affiliation{Max-Planck-Institut f\"ur Physik, F\"ohringer Ring 6, 80805 M\"unchen, Germany}

\begin{abstract}

A system of classical interacting spins can develop collective instabilities which, in the nonlinear regime, mimic the motion of a gyroscopic pendulum. Known as the flavor pendulum, this behavior appears among the collective modes of a dense neutrino plasma after a strong reduction of phase space through symmetry assumptions. It has been identified in homogeneous slow and fast flavor systems and, most recently, in single-wave solutions of the fast system. We explain the reasons for its ubiquitous appearance. We show that a system of three classical spins must always be pendular, or only two in the presence of an external field. Furthermore, such a system always defines a continuum of vectors with time-independent length. If these are identified as interacting spins, they immediately lead to the continuum cases of slow and fast flavor pendula. As another new insight, any of these spins can be chosen as \textit{the} pendulum, periodically exchanging flavor with the rest of the system.

\end{abstract}

\maketitle

\section{Introduction}

Neutrino-neutrino refraction~\cite{Pantaleone:1992eq} leads to collective \hbox{flavor}  evolution~\cite{Samuel:1993uw, Samuel:1995ri, Duan:2006an} that should be included in the neutrino transport in core-collapse supernovae and neutron-star mergers \cite{Duan:2009cd, Duan:2010bg, Tamborra:2020cul, Volpe:2023met, Johns:2025mlm, Raffelt:2025wty}. The mean-field kinetic equations are well known \cite{Dolgov:1980cq, Rudsky, Sigl:1993ctk, Fiorillo:2024fnl, Fiorillo:2024wej}, but elude first-principle implementation because of the small length and time scales compared with the hydrodynamical ones. With the long-term goal of well-founded practical implementations, we have recently developed a complete theoretical framework describing the collective modes of the neutrino plasma, both in its linearized \cite{Fiorillo:2024bzm, Fiorillo:2024uki, Fiorillo:2024pns, Fiorillo:2025ank, Fiorillo:2024dik, Fiorillo:2025zio,Fiorillo:2025gkw,Fiorillo:2025kko} and nonlinear form~\cite{Fiorillo:2024qbl, Fiorillo:2025npi}. However, we here return to a more traditional topic in this field, the flavor pendulum, that has emerged in different incarnations of collective flavor evolution.

The first instance was Samuel's discovery of bimodal coherent behavior in a homogeneous neutrino gas~\cite{Samuel:1995ri}, where the flavor of the ensemble evolved in a surprising form of self-maintained coherence that was in some sense equivalent to two interacting polarization vectors. This early example was, in modern language, a case of \textit{slow} flavor evolution, i.e., driven by the neutrino mass splittings.

Quite generally, in the two-flavor limit, neutrino flavor dynamics, like that of any two-level system, is formally identical to a spin system, often called flavor spin or flavor isospin. In the mean-field approach, these are classical variables, and after a strong reduction of phase space by symmetry assumptions, one is left with an ensemble of interacting classical spins, often represented by polarization vectors $\bP_v$. They depend only on a single remaining phase-space parameter that we call $v$, and represent the flavor density matrix in the form $\varrho_v=\frac{1}{2}\bP_v\cdot{\bm\sigma}$ with ${\bm\sigma}$ a vector of Pauli matrices. 

It was later recognized that the motion of two classical spins interacting with each other and with an external magnetic field $B$ can be mapped onto the dynamics of a mechanical gyroscopic pendulum \cite{Hannestad:2006nj, Duan:2007mv}. Here we show that this correspondence is not merely a curiosity, but rather an unavoidable feature of systems of interacting spins. The question of \textit{why a pendulum\/} has become more pressing with the discovery of additional instances, in fast flavor systems \cite{Johns:2019izj, Padilla-Gay:2021haz, Fiorillo:2023mze}, in which neutrinos are taken to be massless, and, most recently, in the single-wave solution of the fast flavor system \cite{Liu:2025muc, Fiorillo:2026ybk, Fiorillo:2026vfo}.

Besides the pendular dynamics itself, the most peculiar aspect of bimodal coherence is its extension to a large or even continuous collection of spins. Examples are the homogeneous and isotropic multi-energy slow system and the energy-integrated multi-angle fast flavor system. Here, all spins, despite their different coupling strengths to each other or to the external field, move collectively such that they simultaneously return to their initial positions after one pendular swing. This collective behavior is a special case of more general $N$-mode coherent solutions~\cite{Raffelt:2011yb}. Mathematically, such regular behavior derives from the integrability of the underlying Hamiltonian, owing to its large number of so-called Gaudin invariants \cite{Gaudin:1976sv, Pehlivan:2011hp, Raffelt:2011yb, Fiorillo:2023mze}. In contrast, this does not apply to the single-wave system, since its minimal flavor pendulum does not extend to a larger collection of spins.

One key aspect of the pendulum is that it represents the simplest realization of an unstable system, in which all spins initially align, and then spontaneously develop large transverse components. Such runaway behavior can be identified through linear stability analysis \cite{Sawyer:2008zs, Banerjee:2011fj, Izaguirre:2016gsx}, which yields both the complex eigenfrequency and the spectral structure of the unstable perturbation. The pendulum solution is the continuation of this linear instability into the nonlinear regime. However, this continuation as a regular periodic solution is not guaranteed to exist. It does exist for the slow and fast flavor systems, but not for the single-wave system. Therefore, the existence of a pendulum is different from that of bimodal coherence.

We will here show, independently from the traditional notion of Gaudin invariants, why the single-wave flavor pendulum does not extend to more spins, whereas the other systems do.

Another noteworthy aspect of pendular dynamics is the surprising flexibility in how the pendulating radius vector can be defined. Traditionally, this vector is constructed as a certain collective quantity derived from the ensemble of all spins. However, it can also be associated with each individual spin, which can then be seen as periodically exchanging flavor with the rest of the system.

We restrict our discussion to the two-flavor case, which admits a mapping onto spin variables. Three flavors, in contrast, involve eight-dimensional polarization vectors with much more complex dynamics, which have also been studied \cite{Dasgupta:2007ws}. Extending our present results to three flavors is beyond the scope of this work. Nevertheless, we have previously shown that the exact integrability of the two-flavor system, guaranteed by the existence of Gaudin invariants, also persists in the three-flavor case \cite{Fiorillo:2023mze}.

We also do not touch upon the collisional instability~\cite{Johns:2021qby}, driven by a difference in neutrino and antineutrino collision rates \cite{Johns:2021qby, Xiong:2022zqz, Lin:2022dek, Fiorillo:2023ajs,Fiorillo:2025zio}. For a monoenergetic, isotropic neutrino-antineutrino system, it appears that during limited phases of time and for specific classes of instabilities a pendular dynamics may emerge~\cite{Johns:2023xae}, but its nature and origin are certainly quite different from the dissipationless dynamics highlighted here. For the standard fast flavor pendulum, the effect of collisions is primarily to damp the coherent, regular oscillations~\cite{Padilla-Gay:2022wck,Fiorillo:2023ajs}. In all cases considered here, we assume the neutrino free streaming limit and thus the conservation of length of the individual polarization vectors.

To set up the overall framework, we begin in Sec.~\ref{sec:systems} with a definition of the slow, fast, and single-wave systems as abstract ensembles of interacting spins. In Sec.~\ref{sec:gyroscope}, we introduce the classical gyroscopic pendulum, also in its most abstract formulation, and discuss its linearization as well as the exact soliton solution. Next, in Sec.~\ref{sec:Interacting-Spins}, we introduce the explicit mappings of each of the different spin systems on an abstract pendulum. In Sec.~\ref{eq:WhyPendulum}, we finally answer our initial question of why pendular dynamics is so generic. We then turn, in Sec.~\ref{sec:continuum}, to the question of when and how such minimal systems of a few spins can be extended to a continuum, before turning to a summary~in~Sec.~\ref{sec:summary}.

\section{Systems of Interacting Spins}

\label{sec:systems}

In the strongly phase-space reduced systems under discussion, the scalar phase-space index $v$ plays a different role in different cases. In the homogeneous or single-angle slow flavor system, it would be interpreted as $v=\omega/\mu$ with $\omega=\delta m^2/2E$ the vacuum oscillation frequency and \smash{$\mu=\sqrt{2}\GF n_\nu$} an effective neutrino-neutrino interaction energy, and $-\infty<v<+\infty$. On the other hand, for axisymmetric fast flavor systems, where vacuum oscillations are neglected, $v=\cos\theta$ is the projection of the neutrino velocity on the symmetry axis with $-1\leq v\leq+1$. However, as we are here interested in the abstract structure of the equations, the concrete physical interpretation or range of $v$ are not important.

For such systems, the phase-space reduced equations of motion (EoMs) are, after absorbing the neutrino-neutrino interaction strength in the units of space and time,
\begin{equation}\label{eq:precession}
    \dot\bP_v=\bH_v\times\bP_v.
\end{equation}
For vacuum oscillations, the Hamiltonian vector would simply be $\bH_v=v\bB$ with an external unit vector $\bB$ pointing in the mass direction in flavor space. In addition, there is a term arising from neutrino-matter refraction that we always leave out here.

Including neutrino-neutrino refraction, three different cases of interacting spin systems have appeared in the literature. To formulate them compactly, we use the moment notation
\begin{equation}\label{eq:moments}
   \bP_n=\int dv\,v^n\,\bP_v, 
\end{equation}
with a slight notational ambiguity in that, for instance, $\bP_0$ means the zeroth moment, not $\bP_v$ at $v=0$. 

Traditional slow flavor evolution for a single-angle or isotropic system is engendered by~\cite{Samuel:1995ri, Duan:2006an, Hannestad:2006nj, Raffelt:2011yb}
\begin{equation}\label{eq:EoM-slow}
    \bH_v=\bP_0+v\bB.
\end{equation}
The fast system, instead, is characterized by \cite{Johns:2019izj, Padilla-Gay:2021haz, Fiorillo:2023mze}
\begin{equation}\label{eq:EoM-fast}
    \bH_v=\bP_0-v\bP_1,
\end{equation}
without an external vector, only depending on neutrino-neutrino refraction. The total angular momentum $\bP_0$, defined by initial conditions, plays the role of a conserved reference vector. Finally, the recently discovered single-wave (SW) system, is spawned by \cite{Liu:2025muc, Fiorillo:2026ybk, Fiorillo:2026vfo}
\begin{equation}\label{eq:EoM-SW}
    \bH_v=\bP_0-v(\bP_1+\bK),
\end{equation}
where $\bK$ is an external vector. It represents the wavenumber along the symmetry direction of an axisymmetric fast flavor system, and in flavor space points in the weak-interaction direction.

Each of these EoMs descends from a classical Hamiltonian if we interpret each $\bP_v$ as a variable obeying the angular-momentum Poisson brackets. The slow, fast, and SW Hamiltonians are
\begin{eqnarray}
    H_{\rm slow}&=&\half\bP_0^2+\bB\cdot\bP_1,
    \\
    H_{\rm fast}&=&\half(\bP_0^2-\bP_1^2),
    \\
    H_{\rm SW}&=&\half(\bP_0^2-\bP_1^2)-\bK\cdot\bP_1,
\end{eqnarray}
which also imply the existence of a conserved energy in all of these examples.

In the context of flavor conversion, the key property of such spin-spin system, and of course the less constrained kinetic equations, is the possible appearance of unstable collective forms of evolution. In particular, if initially all spins are aligned along the $z$-direction in flavor space, this configuration can be an unstable fixed point, and a slight perturbation allows the $\bP_v$ to develop large transverse components. Therefore, in a dense neutrino plasma, flavor coherence can arise even when vacuum oscillations are suppressed by matter effects. In the linearized problem of small deviations of the $\bP_v$ relative to the $z$-direction, such instabilities can be found by solving a well-known dispersion relation.

One surprising property of such instabilities is that in a system of interacting spins, the linear solution can continue into the nonlinear regime such that all spins collectively move away from the $z$-direction and return to it in a completely ordered manner, with different maximum amplitudes. In other words, there is a characteristic $v$-dependent pattern of oscillation amplitudes, and a characteristic form of time evolution, determined by the underlying dynamics of a gyroscopic pendulum. These are two somewhat separate phenomena.

The characteristic pendulum-like dynamics exists for a minimum of two $\bP_v$ modes in the slow and SW systems, each of which has an external vector $\bB$ or $\bK$, defining an ``external magnetic field'' that affects the spins besides their mutual interaction. Or we need a minimum of three modes in the fast system, that lacks an external field, and therefore, if there were only two modes, these would simply precess around each other without pendular dynamics. In all cases, we need three vector variables, one of which is conserved, and the other two are dynamical.

In the slow and fast system, but not the SW system, the pendulum motion can be extended to a continuum of $\bP_v$, each of which evolves as a linear superposition of the original three (two dynamical plus one conserved) and yet fulfill the respective set of precession equations of Eqs.~\eqref{eq:EoM-slow} or \eqref{eq:EoM-fast}. The possibility of such bimodal or two-mode coherent solutions follows from the fact that the slow and fast Hamiltonians are technically integrable (they possess so-called Gaudin invariants), whereas the SW Hamiltonian is not integrable.

Taking the abstract EoMs of Eqs.~\eqref{eq:EoM-slow}--\eqref{eq:EoM-SW} as a starting point, and without worrying about their physical interpretation, we will first address the question of why minimal versions of these equations \textit{must} evolve like pendula, i.e., what are the common characteristics that make pendulum dynamics inevitable. Already this simple question leads to a surprisingly general answer, and shows the flexibility of which variables are actually interpreted to play the role of the different pendulum variables of radius vector $\bR$, angular momentum $\bJ$, and gravitational field $\bG$.

The second line of investigation is about systems of many (or a continuum of) $v$-modes and how they generally connect to the two or three mode cases. Moreover, we will see that in such systems, any of the $\bP_v$, or a certain collective combination, can play the role of the quantity $\bR$ that pendulates, once more showing a great flexibility of interpretation. Using a specific $\bP_v$ instead of the traditional collective variable is especially advantageous for weak instabilities, when only a narrow range of $v$-modes strongly moves, and it is most natural to use the exactly resonant one as \textit{the\/} pendulum.

\section{Gyroscopic Pendulum}

\label{sec:gyroscope}

\subsection{Definition}

A gyroscopic pendulum, or heavy top, or Lagrange top, as studied in classical mechanics textbooks \cite{Klein-Sommerfeld:2008}, consists of three vector variables $\bG$, $\bR$, and $\bJ$. After absorbing all dimensional parameters, such as the moment of inertia, into these variables and into the units of time, these variable obey the EoMs
\begin{equation}\label{eq:pendulum}
    \dot{\bG}=0,\quad\dot{\bR}=\bJ\times\bR,\quad\dot{\bJ}=\bG\times \bR.
\end{equation}
In the mechanical case, the vector $\bG$ represents gravity and is conserved, $\bR$ is the radius vector connecting the point of support with the center of mass and has conserved length, and $\bJ$ is the total angular momentum. It engenders a differential rotation on $\bR$, and in turn, is modified by a torque exerted by gravity pulling at the center of mass. The dynamics is the same if we scale $\bR$ and $\bG$ such that $\bR\cdot\bG$ is the same. Therefore, one could choose one of them to be a unit vector.

Besides $\bG$ and $\bR^2$, other conserved quantities are the projection of $\bJ$ on $\bG$, the latter defining the $z$-direction, so $J_z$ is conserved. Moreover, the projection of $\bJ$ on $\bR$ is also conserved, called the spin $\bS$ of the gyroscope. Physically, the gravitational force acting on the symmetry axis cannot exert a torque to modify $\bS$. Finally, energy is conserved and indeed, the EoMs descend from the classical Hamiltonian
\begin{equation}
    H=\frac{1}{2}\bJ^2+\bG\cdot\bR,
\end{equation}
where $\bG\cdot\bR$ is the potential energy. Overall, we are left with two dynamical variables, that are often taken to be the polar angles of~$\bR$ and specifically $\vartheta$, defined as the angle between $\bG$ and $\bR$, and an azimuth angle $\varphi$. 

\subsection{Linearization}\label{sec:linearization}

In the neutrino context, we are mostly interested in configurations with an initially vertical pendulum, collinear with $\bG$, that defines the $z$-direction. Most interesting is the inverted position, allowing a small perturbation to make the pendulum tilt and the transverse component of $\bR$ to grow exponentially. To quantify this behavior, we may initially use the linearized equations that apply when the transverse components of both $\bR$ and $\bJ$ are small.

It is convenient to split all vectors $\bP$ into their $z$-component $P_z$ and transverse part $P_\perp=P_x+i P_y$. Besides $\dot G_z=0$ and $G_\perp=0$, the full pendulum equations are
\begin{subequations}
    \begin{eqnarray}
    \dot R_z\,&=&\frac{i}{2}\bigl[J_\perp R_\perp^*-J_\perp^*R_\perp\bigr],\\
    \dot R_\perp&=&\,i\,\bigl[J_z R_\perp-R_z J_\perp\bigr],\\
    \dot J_z\,&=&0,\\[0.3ex]
    \dot J_\perp&=&\,i\,G_z R_\perp.
    \end{eqnarray}
\end{subequations}
Both $G_z$ and $J_z$ are conserved, and in the linear regime, we can take $R_z$ to be fixed as well because initially it changes only to second order. The linear EoMs are
\begin{equation}
    \partial_t\begin{pmatrix}R_\perp\\ J_\perp\end{pmatrix}=i
    \begin{pmatrix}J_z&-R_z\\ G_z&0\end{pmatrix}
    \begin{pmatrix}R_\perp\\ J_\perp\end{pmatrix},
\end{equation}
and with a time variation assumed to be $e^{-i\Omega t}$, the eigenfrequencies are
\begin{equation}
    \Omega=\frac{J_z}{2}\pm \sqrt{\left(\frac{J_z}{2}\right)^2-G_z R_z}
    \,.
\end{equation}
If the pendulum begins in a vertical position, the initial $\bJ$ is identical with the spin $\bS$, so the initial and conserved $J_z$ is the same as the spin $S$. Moreover, the initial potential $\Phi=G_zR_z|_{\rm ini}$ can be positive or negative, depending on whether the pendulum is initially in the downward position ($\Phi<0$) or the inverted one ($\Phi>0$).

For vanishing spin, $S=0$, and with a downward initial orientation, the pendulum executes small-angle harmonic oscillations with the natural pendulum frequency
\begin{equation}
    \lambda=\sqrt{|G_z R_z|_{\rm ini}}
    =\sqrt{|\Phi|}.
\end{equation}
If instead it begins with an upright orientation, the transverse component grows or shrinks as $e^{\pm\lambda t}$.

An instability requires $\Phi>0$ and $(S/2)^2<\Phi$. In other words, it requires the pendulum to begin in the upright orientation, and it must not spin too fast, or else it will be in the sleeping-top configuration, an inverted gyroscope spinning so fast it cannot fall. In the unstable case, $\Omega=\Omega_R+i\gamma$ with
\begin{equation}\label{eq:eigenfrequency-pendulum}
  \Omega_R=S/2
   \quad\text{and}\quad
   \gamma=\sqrt{(S/2)^2-\Phi}.
\end{equation}
Sometimes it is useful to introduce the spin parameter
\begin{equation}
    \sigma=\frac{S}{2\lambda}=\frac{S}{2\sqrt{|\Phi|}}.
\end{equation}
For an upright pendulum ($\Phi>0$), an instability occurs for $0<\sigma^2<1$. 

\subsection{Explicit solution}
\label{sec:soliton}

The pendulum equations \eqref{eq:pendulum} can be solved analytically beyond the linear regime as shown in the neutrino context in Refs.~\cite{Raffelt:2011yb, Padilla-Gay:2021haz, Fiorillo:2023mze, Fiorillo:2023hlk}. Beginning in the upright position and no initial kick, the polar coordinates obey
\begin{subequations}\label{eq:pendulum-solution}
    \begin{eqnarray}
        \dot\varphi&=&\frac{2\lambda\sigma}{1+c_\vartheta}=\frac{S}{1+c_\vartheta},
  \\[1ex]
  \dot c_\vartheta^2&=&2\lambda^2(1-c_\vartheta)^2(1+c_\vartheta-2\sigma^2),
    \end{eqnarray}
\end{subequations}
where $c_\vartheta=\cos\vartheta$. For $\sigma=0$, the motion is that of a plane pendulum with only $\vartheta$ moving. On the other extreme, for $|\sigma|\ge1$, the right-hand side of the second equation is negative for any $c_\vartheta$ so that there is no solution other than $c_\vartheta=1$ and $\dot c_\vartheta=0$, the sleeping-top configuration mentioned earlier.

In the unstable configuration ($0<|\sigma|<1$), and when initially $c_\vartheta$ is very close to one, the initial exponential growth phase for $\vartheta$ lasts arbitrarily long, depending on the initially chosen small angle. The limiting solution for a perfectly upright pendulum at $t\to -\infty$ could be called an instanton, but we have preferred the terminology of a temporal soliton~\cite{Fiorillo:2023hlk} since instanton in quantum field theory refers to a transition happening in complex, rather than real, time. Either way, the motion starts in the infinite past and the pendulum returns to the upright position in the infinite future. With $t=0$ as the instant of maximum excursion, the explicit solution is \cite{Fiorillo:2023hlk}
\begin{eqnarray}\label{eq:instanton}
  c_\vartheta&=&1-2(1-\sigma^2)\,{\rm sech}^2\left(\sqrt{1-\sigma^2}\lambda t\right)
  \nonumber\\
  &=&1-2\,\frac{\gamma^2}{\gamma^2+\Omega_R^2}
  \,{\rm sech}^2\left(\gamma t\right),
\end{eqnarray}
where ${\rm sech}(x)=2/(e^{-x}+e^{x})$. If the system actually begins with a small but nonzero initial angle $\vartheta_0$, a full swing takes a finite amount of time, leading to a periodic motion, of which the temporal soliton is a limiting case.

\section{Interacting Spins as a Pendulum}

\label{sec:Interacting-Spins}

\subsection{Minimal slow system}

Two classical spins interacting with an external magnetic field and with each other were the first case for the mapping to pendulum dynamics \cite{Hannestad:2006nj}. The slow EoMs of Eq.~\eqref{eq:EoM-slow} for a set of two spins with frequencies $v_1$ and $v_2$ represent a pendulum with the identification
\begin{subequations}\label{eq:slow-pendulum-mapping}
\begin{eqnarray}
    \bG&=&\bB,\\
    \bR&=&v_1v_2\bB+v_1\bP_{v_1}+v_2\bP_{v_2}\\
    \bJ&=&(v_1+v_2)\bB+\bP_{v_1}+\bP_{v_2}.
\end{eqnarray}    
\end{subequations}
This identification is unique except for an arbitrary scaling factor for $\bG$ and $\bR$ such that $\bG\cdot\bR$ remains the same. Using the moment notation of Eq.~\eqref{eq:moments}, these equations are more compactly $\bR=v_1v_2\bB+\bP_1$ and $\bJ=(v_1+v_2)\bB+\bP_0$. One can verify that the new variables fulfill the pendulum Eqs.~\eqref{eq:pendulum} if the original variables fulfill the slow flavor EoMs \eqref{eq:EoM-slow} and one can verify that $\bR^2$ is indeed conserved.

\subsection{Rotating frame}

One can also study the pendulum dynamics in a frame corotating around $\bB$ with frequency $\vc$. This approach is especially motivated when the initial spins are aligned with the $z$-direction. In a rotating frame, the evolution is not the same because the initial conditions also need to be taken in the new frame. On the other hand, beginning in an unstable fixed point and simply following the unstable solution after a small disturbance amounts to the same physics, and notably to the same $z$-projection of the solution.

In the new frame, the spins labeled as $\bP_{v_1}$ and $\bP_{v_2}$ have the vacuum precession frequencies $v_1'=v_1-\vc$ and $v_2'=v_2-\vc$, i.e., Eq.~\eqref{eq:EoM-slow} is 
\begin{equation}\label{eq:slow-c}
    \bH_v^{\rm c}=\bP_0+(v-\vc)\,\bB.
\end{equation}
In this case, the explicit mapping of Eq.~\eqref{eq:slow-pendulum-mapping} to a pendulum  is the same with $v_1\to v_1-\vc$ and $v_2\to v_2-\vc$. Thus, there is a continuum of ways to formulate the pendulum dynamics, but there are three special cases in addition to the one in the original frame.

The first is the choice made in the early pendulum papers, with  $\vc=(v_1+v_2)/2$ so that the two spins precess with the same frequency in opposite directions, implying that $\bJ=\bP_0$, i.e., the total angular momentum of the pendulum coincides with that of the original system, but the expression for $\bR$ remains complicated.

The other special choices are $\vc=v_1$ or $v_2$, i.e., the observer corotates with one of the vacuum frequencies. In this case, it is one of the original spins playing the role of $\bR$, and concentrating on $\vc=v_2$, it is $\bR=(v_1-v_2)\bP_{v_1}$. We could also use $\bR=\bP_{v_1}$ together with $\bG=\bB/(v_1-v_2)$ such that $\bG\cdot\bR$ remains the same. Either way, the angular momentum is $\bJ=(v_1-v_2)\bB+\bP_{v_1}+\bP_{v_2}$.

Indeed, the original precession Eq.~\eqref{eq:precession} already suggests the identification of $\bP_{v_1}$ (and also $\bP_{v_2}$) playing the role of $\bR$ in that the length of $\bP_{v_1}$ is conserved and there is a constant vector $\bB$ to play the role of $\bG$. However, one easily verifies that $\bH_{v_1}$, while engendering a differential rotation for $\bP_{v_1}$, cannot be interpreted as $\bJ$ as it would not fulfill the third pendulum equation. This mismatch can then be fixed by going to a rotating frame.

\subsection{Same instability?}

In different formulations of the same pendulum dynamics, the solution for every $\bP_{v}(t)$ is the same, except for an additional rotation around $\bB$. However, the explicit pendulum parameters, notably the natural pendulum frequency $\lambda$ and spin $S$, are different in each case. 

However, what should be the same is the initial growth rate $\gamma$, the imaginary part of the eigenfrequency for the linearized system Eq.~\eqref{eq:eigenfrequency-pendulum}. In the neutrino context, we usually assume the polarization vectors to be initially aligned with the $z$-direction, and we usually denote the $z$-components with $D_v=P_v^z$ and the initial values as $D_v^0=P_v^z(t=0)$. With such initial conditions, the initial pendulum potential is
\begin{equation}
    \Phi=v_1^{\rm c}v_2^{\rm c}+v_1^{\rm c}D_{v_1}^0+v_2^{\rm c}D_{v_2}^0,
\end{equation}
where $v_1^{\rm c}=v_1-\vc$ etc., recalling that $\bB$ is a unit vector defining the $z$-direction. The spin, the projection of the initial $\bJ$ on $\bG=\bB$, is
\begin{equation}
    S=v_1^{\rm c}+v_2^{\rm c}+D_{v_1}^0+D_{v_2}^0,
\end{equation}
implying with Eq.~\eqref{eq:eigenfrequency-pendulum}
\begin{equation}
    \gamma=\sqrt{(D_1^0+v_1v_2)-(D_0^0+v_1+v_2)^2/4}\,,
\end{equation}
where $\vc$ has indeed dropped out.

\subsection{Minimal fast system}

For the fast system, one needs a minimum of three spins, as explained earlier. In this case, it is easier to express the pendulum parameters immediately in terms of the first three moments $\bP_n$ ($n=0,1,2$) that are equivalent to the three spins $\bP_{v_i}$ ($i=1,2,3$). Here the mapping to a pendulum was found to be \cite{Padilla-Gay:2021haz}
\begin{subequations}
    \begin{eqnarray}
        \bG&=&v_1v_2v_3\,\bP_0,\\
        \bR&=&\bP_1,\\
        \bJ&=&\bP_2-(v_1+v_2+v_3)\,\bP_1.
    \end{eqnarray}
\end{subequations}
To obtain such a pendulum, none of the $v_i$ must vanish. As always, one can rescale $\bG$ and $\bR$ such that $\bG\cdot\bR$ remains the same, but otherwise the mapping is unique.

Once more, one can view the system from the perspective of an observer corotating around $\bP_0$ such that any of $\bP_{v_i}$ ($i=1$, 2 or 3) will play the role of $\bR$. To show this explicitly, we rewrite the EoMs of the fast Hamiltonian in Eq.~\eqref{eq:EoM-fast} in terms of $\bP_{v_1}$, $\bP_{v_2}$, and $\bP_0$ as
\begin{equation}\label{eq:EoM-Pv1}
    \dot\bP_{v_1}=-v_1v_3\bP_0\times\bP_{v_1}+v_1(v_3-v_2)\bP_{v_2}\times\bP_{v_1}
\end{equation}
and analogous for $\bP_{v_2}$ with indices exchanged. Going to a corotating frame around the conserved $\bP_0$ with angular velocity $-v_2 v_3|\bP_0|$, Eq.~\eqref{eq:EoM-Pv1} becomes $\dot\bP_{v_1}=\bJ\times \bP_{v_1}$ with $\bJ=v_3(v_2-v_1)\bP_{v_1}+v_1(v_3-v_2)\bP_{v_2}+v_2(v_3-v_1)\bP_{v_1}$. The last term is chosen such that the angular momentum obeys the final equation $\dot\bJ=v_2 v_3(v_3-v_1)(v_2-v_1)\bP_0\times\bP_{v_1}$, completing the pendulum correspondence with $\bR=\bP_{v_1}$ as the radius, $\bJ$ as the angular momentum, and $\bG=v_2 v_3(v_3-v_1)(v_2-v_1)\bP_0$ as the gravity vector. The same construction can be realized for any of the three spins. 

Once more, while the various pendulum parameters are different, the instability properties are the same. Each formulation yields the same solutions for the individual functions $\bP_{v_i}(t)$, except for the different corotations.

\subsection{Minimal single-wave system}

For the SW system of Eq.~\eqref{eq:EoM-SW}, a minimum of two dynamical modes $\bP_{v_1}$ and $\bP_{v_2}$ in addition to the external $\bK$ are required to form a pendulum. The corresponding mapping is found to be \cite{Fiorillo:2026ybk} 
\begin{subequations}\label{eq:SW_first_formulation}
  \begin{eqnarray}
      \bG&=&\bK,\\
      \bR&=&v_1v_2\bK-(1-v_1v_2)\bP_1,\\
      \bJ&=&-(v_1+v_2)\bK+(1-v_1v_2)\bP_0.
  \end{eqnarray}  
\end{subequations}
Once more, the pendulum parameters are unique up to a rescaling of $\bR$ and $\bG$ such that $\bG\cdot\bR$ is fixed. 

One can also go to frames corotating around $\bK$ such that one of the original spins plays the role of $\bR$. Considering, in particular, $\bP_{v_1}$, the required corotation is with frequency $\vc=v_2 K$. In this case, the mapping is \cite{Fiorillo:2026ybk}
\begin{subequations}
  \begin{eqnarray}
      \bG&=&(1-v_1v_2)(v_2-v_1)\bK,\\
      \bR&=&\bP_1,\\
      \bJ&=&(v_2-v_1)\bK+(1-v_1v_2)\bP_0.
  \end{eqnarray}  
\end{subequations}
For the second beam, analogous expressions obtain by $1\leftrightarrow2$. Once more, any of these pendulum formulations yields different $\lambda$ and $S$, but the same growth rate $\gamma$ and they describe the same physical motion \cite{Fiorillo:2026ybk}.

\section{When Three Make a Pendulum}

\label{eq:WhyPendulum}

The emergence of simple pendular dynamics in these different systems of interacting spins was surprising in each new case, but of course, there must be a general principle behind it. All of these cases have two crucial common ingredients. One is a conserved vector, specifically $\bB$, $\bP_0$, and $\bK$ in the slow, fast, and SW cases, respectively. The other ingredient is that the motion of each spin $\bP_v$ is an instantaneous precession, preserving its length. This property justifies the terminology of ``spin'' rather than that of a general angular momentum that might suffer a torque and change its magnitude. 

These two ingredients are enough to explain the equivalence to a pendulum. Consider a triplet of vector variables $\{\bR, \bQ, \bG\}$, of which $\bG$ is actually conserved and $\bR$ has conserved length. The most general evolution equations for $\bR$ and $\bQ$ must have the structure
\begin{subequations}
    \begin{eqnarray}
        \dot{\bR}&=&(\alpha \bG+\beta\bQ)\times\bR,
        \\
        \dot{\bQ}&=&\gamma \bG\times \bQ+\eta \bG\times\bR+\xi\bR\times \bQ.
    \end{eqnarray}
\end{subequations}
These can always be reduced to the pendulum equations. Let us first go to a corotating frame around $\bG$ with angular velocity $b$. (We will denote by Latin letters the quantities that are adjustable parameters of our procedure and are not fixed by the original equations.) The EoMs are then modified to
\begin{subequations}
    \begin{eqnarray}
        \dot{\bR}&=&[(\alpha-b) \bG+\beta\bQ]\times\bR,
        \\
        \dot{\bQ}&=&(\gamma-b) \bG\times \bQ+\eta\bG\times\bR+\xi\bR\times \bQ.
    \end{eqnarray}
\end{subequations}
Since $\bR$ is what plays the role of the pendulum radius vector, the angular momentum must be the operator in the first line, except for the possible addition of an unknown multiple of $\bR$, so that $\bJ=(\alpha-b)\bG+\beta\bQ+c\bR$. The rate of change of this vector, after replacing the time derivatives, becomes
\begin{eqnarray}
    \dot{\bJ}&=&\bG\times\bR\,
    \bigl[\beta\delta-(\gamma-b)c+\xi(\alpha-b)\bigr]
    \nonumber\\[1ex]
    &&{}+\bJ\times\bR(c-\xi)+\bG\times\bJ(\gamma-b).
\end{eqnarray}
For this to be the equation of a gyroscopic pendulum, the right-hand side must be of the form $\dot{\bJ}=g\bG\times\bR$. Comparing coefficients implies $c=\xi$, $b=\gamma$, and $g=\beta\eta+\xi\alpha-\xi\gamma$. Finally, we can absorb $g$ in the definition of $\bR$ or $\bG$. 

In conclusion, from the perspective of a corotating observer, the dynamics of the original triplet $\{\bR, \bQ,\bG\}$ is equivalent to a gyroscopic pendulum. All of the slow, fast, and SW systems are of this form so that their minimal realizations, consisting each time of a total of three vector variables, inevitably must evolve like gyroscopic pendula. 

The general result is that any triad of vectors, of which at least two have conserved length, engaged in a mutual interaction, necessarily is identical to a pendulum. Actually, this result can be generalized even further. Given a triad of vectors, it suffices that only one of them has a conserved length to make the dynamics pendular. If one vector $\bG$ has conserved length, one can always choose a frame in which it is conserved. The other two, say, $\bA$ and $\bB$, obey the most general EoMs
\begin{subequations}
\begin{eqnarray}
    \dot\bA&=&\mu\bB\times\bA+\lambda\bG\times\bA+\theta\bG\times\bB,\\
    \dot\bB&=&\nu\bA\times\bB+\alpha\bG\times\bA+\beta\bG\times\bB.
\end{eqnarray}    
\end{subequations}
It is now easy to verify that, even if we have not postulated any other fixed-length vector, there is always a combination $\bA+c\bB$ such that its length is conserved. Hence, for any three vectors engaged in a mutual interaction, it suffices that one of them has conserved length to ensure that the system is a pendulum.

\section{Continuum of Spins}

\label{sec:continuum}

One surprising feature of an ensemble of classical spins, $\bP_v$,  with the different physical interpretations of the parameter $v$ discussed earlier, is that many spins, or even a continuum, can show perfectly regular behavior over the entire $v$-spectrum. In particular, all $\bP_v$ may simultaneously follow the pendulum dynamics, meaning that all return to vertical positions at the same instant if they started there initially. We now explore different perspectives on this bimodal coherent behavior. This phenomenon is a nontrivial extension of the pendular behavior of the minimal systems studied in the previous sections and does not apply to the SW case.

\subsection{Lax vectors}

To appreciate the reason for such coherent behavior, we may consider a minimal flavor pendulum and add an additional spin that we call $\bL_u$ with a length so small that it does not impact the motion of the original spins. It must follow the original precession equation~\eqref{eq:precession} in the form $\dot\bL_u=\bH_u\times\bL_u$. Its motion will be coherent with that of the original minimal set of spins if its time evolution is a linear superposition of that of the original ones. To be specific, we consider the slow system and one finds that, up to a global normalization factor,
\begin{equation}\label{eq:lax-slow}
    \bL_u=(u-v_1)(u-v_2)\bB+(u-v_2)\bP_{v_1}+(u-v_1)\bP_{v_2}
\end{equation}
is a unique solution for any choice of $u$, as one can verify by explicit replacement. The new spin thus constructed is called a Lax vector of the original system.

As a consequence, a test spin $\bL_u$, which couples to $\bB$ with strength $u$, and beginning in the vertical position, will evolve in the superposition of Eq.~\eqref{eq:lax-slow}, which can be expressed in terms of the pendulum variables engendered by the self-consistent motion of $\bP_{v_1}$ and $\bP_{v_2}$. 

Of course, if one were to begin with $N$ original spins that, depending on their properties and initial conditions, can evolve in more complicated ways than a pendulum, also allow for the construction of analogous Lax vectors \cite{Raffelt:2011yb}. This works out so magically because the Hamiltonian is integrable based on the existence of the so-called Gaudin invariants. However, for the purpose of studying the flavor pendulum, such more general cases are not important. It is enough to show that a test spin follows the original pendulum through the construction of Eq.~\eqref{eq:lax-slow}.

For the fast system, beginning with the minimal three-spin ensemble, the analogous construction is \cite{Fiorillo:2023mze}
\begin{equation}
    \bL_u=\left(\sum_{i=1}^3 \frac{v_i\bP_{v_i}}{u-v_i}\right)
    \prod_{i=1}^3(u-v_i).
\end{equation}
Once more, an additional test spin moves coherently with the others in the form of this superposition.

For the SW system, on the other hand, there is no general linear superposition of the original spins together with the conserved vector $\bK$ to solve the original EoMs. A test spin generally does not evolve in the form of such a superposition. In other words, the SW Hamiltonian is not integrable, it does not possess Gaudin invariants~\cite{Fiorillo:2026ybk}.

For the slow and fast systems, one can finally consider a set of $n$ discrete polarization vectors, or even a continuum, that evolve in the field of the others. Each one follows a pendulum, and in turn, their collective $\bP_0$ is the sum of the individual spins. For this to happen, they need to fulfill a self-consistency condition that has been frequently discussed in the literature. The parameters of the underlying pendulum follows from the parameters $\Omega=\Omega_R+i\gamma$ that are implied by the usual linear stability analysis.

Even for the SW case, one can perform a linear stability analysis and, if an unstable mode exists, find $\Omega=\Omega_R+i\gamma$, the would-be pendulum parameters. However, in this case the linear solution does not continue coherently into the nonlinear regime, there will not be an underlying pendulum solution, and the spins do not return to their original vertical position after some period.

\subsection{One pendulum spawns many}

The existence or nonexistence of a continuous range of spins collectively acting as a flavor pendulum can also be addressed directly from the perspective of the flavor pendulum, without looking at their specific minimal realizations and concomitant Lax vectors.

A system of three variables $\{\bR, \bJ, \bG\}$, fulfilling the pendulum equations, immediately implies a continuum of other three-vector pendula. We first note that there is a one-parameter family of vectors with conserved length, arising as a linear superposition of the original set, of the unique form
\begin{equation}\label{eq:Multi-Lu}
    \bL_u=\bG+u\bJ+u^2\bR.
\end{equation}
For any real $u$, the length of $\bL_u$ is conserved, as one can show by considering $\partial_t\bL_u^2=2\bL_u\cdot\dot\bL_u$ and replacing the definition and the pendulum EoMs. The EoM for the new vector is explicitly
\begin{equation}\label{eq:EoM-Lu}
    \dot{\bL}_u=-u\bR\times \bL_u.
\end{equation}
In the limits $u\to 0$ and $u\to\infty$, the new variable $\bL_u$ tends to the two fundamental vectors of conserved length, namely $\bG$ and $\bR$ respectively. Since the length of $\bL_u$ is arbitrary, we could divide the right-hand side of Eq.~\eqref{eq:Multi-Lu} by $u$, making the expression more symmetric between $\bG$ and $\bR$.

Following the logic of Sec.~\ref{eq:WhyPendulum}, the length-conserved $\bL_u$, together with $\bJ$ and the conserved $\bG$, must form a new pendulum in a certain frame rotating around $\bG$. To see this explicitly, we go to a corotating frame around $\bG$ with frequency $-G/u$, so that $d/dt\to d/dt-(1/u)\bG\,\times$. In this case, from Eq.~\eqref{eq:EoM-Lu} follows that in the new frame, $\dot{\bL}_u=[-u\bR+(1/u)\bG]\times\bL_u$. However, to achieve the pendulum form $\dot{\bL}_u=\bJ_u\times\bL_u$, we need to add a suitable multiple of $\bL_u$ and find that the appropriate form is
\begin{equation}
    \bJ_u=(\bG+\bL_u)/u-u\bR=2\bG/u+\bJ,
\end{equation}
which indeed obeys $\dot\bJ_u=\bG_u\times\bL_u$ with $\bG_u=\bG/u^2$. (Notice that replacing the expressions for $\bG_u$, $\bL_u$, and $\bJ_u$ leads to
$\dot\bJ=\bG\times\bJ/u+\bG\times \bR$, which is the correct pendulum equation for $\bJ$ in the rotating frame.)
It is then intriguing that \textit{any} gyroscopic pendulum implies the existence of a continuous set of vectors which also behave as gyroscopic pendula in different frames. 

Among the whole continuum of possible pendula, identified by their radius vector $\bL_u$, there is a very special one which has zero spin. For an unstable configuration in which initially $\bR$ and $\bJ$ are parallel to $\bG$, this happens when $2G_z+uJ_z=0$, where $J_z$ and $G_z$ are the unperturbed initial components as in Sec.~\ref{sec:linearization}. As we see from Eq.~\eqref{eq:instanton}, a pendulum with zero spin flips over completely in a full swing. The conclusion is that among the continuum of $\bL_u$, there is always one that fully flips over as the instability proceeds.

Our general study of the pendulum gives us a description that is as generic as possible: a pendulum is a system of three mutually precessing vectors, of which two have a fixed length, and one of these two is conserved in some frame. There is a certain arbitrariness in how the conserved vectors are chosen: for example, if we go to a frame rotating around $\bJ$, we can change the equations of motion to $\dot{\bR}=0$ and $\dot{\bG}=-\bJ\times \bG$, so that $\bR$ and $\bG$ exchange roles, becoming gravity and position respectively. So in general in a pendulum any combination of $\bR$ and $\bG$ can be chosen as the pendulum vector, and the orthogonal position will be the gravity vector. 

Historically, how the pendulum came to appear in the flavor conversion problem was through an exact set of equations---the kinetic equations of a homogeneous neutrino plasma---which were realized to possess the special Gaudin conservation laws and therefore admit pendular solutions. However, with our renewed understanding of the pendulum in general, we can ask the question in reverse: under what conditions can a continuous system of polarization vectors exhibit pendular dynamics? The $\bL_u$, in a frame generically corotating around $\bG$ with frequency $a$, satisfy the equations
\begin{equation}\label{eq:precession_L_vectors}
    \dot{\bL}_u=(a\bG-u\bR)\times\bL_u.
\end{equation}
These equations are characterized, in the most generic terms, as a family of L-vectors precessing around a linear combination of the two fixed-length vectors $\bR$ and $\bG$, of which one is conserved; which one depends on the chosen frame. The combination must be linear also in the parameter $u$ describing the family. 

This approach allows us to understand why the slow and fast spin systems admit pendular solutions, while the SW system in general does not. The fast flavor EoM $\dot{\bP}_v=(\bP_0-v\bP_1)\times \bP_v$ conserve $\bP_0$ and the length of $\bP_1$, so they are evidently of the form Eq.~\eqref{eq:precession_L_vectors}. The slow flavor EoMs $\dot{\bP}_v=(v\bB+\bP_0)\times \bP_v$ possess $\bB$ as a fixed vector and $\bP_0$ as one with conserved length, so they are also of the form of Eq.~\eqref{eq:precession_L_vectors}. 

On the other hand, the SW system with the EoM $\dot{\bP}_v=[\bP_0-v(\bP_1+\bK)]\times \bP_v$ has the conserved vector $\bK$, but neither $\bP_0$ nor $\bP_1$, nor any of their combinations, have fixed length. Therefore, generally this system cannot admit pendulum solutions---unless, of course, it is composed of only two polarization vectors, in which case it contains exactly three vectors with the right properties. The conclusion is the same as that following from the absence of Gaudin invariants.

For the fast and slow flavor pendulum, the identification of the $\bP_v$ of the continuous system with the continuous set of $\bL_u$ generated from the pendulum allows us to draw one more conclusion. Since one of the $\bL_u$ is certain to flip over completely in the pendular instability, it follows that among the polarization vectors one is certain to flip over completely, a spinless pendulum. For the fast system, for which $v$ is restricted to lie in the interval $[-1,+1]$, this special spinless polarization vector may turn out to be superluminal ($|v|>1$). If it is subluminal ($|v|\leq1$), however, there is a physical neutrino mode changing completely its flavor in the development of the instability. In the limit of a weak instability, with $\gamma\to 0$, this special spinless mode must coincide with the resonant neutrinos: we have shown this explicitly in Ref.~\cite{Fiorillo:2026ybk}. 

Indeed, we have also shown in Ref.~\cite{Fiorillo:2026vfo} that this conclusion is somewhat more general: any weak instability, regardless of whether it maps to an exact pendulum, possesses an approximate pendular behavior with a resonant neutrino mode flipping completely its flavor under the action of the collective field produced by all the other neutrino modes. This observation establishes a direct connection between the pendular dynamics, which is exact under the special circumstances discussed here, and the generic resonant mechanism at work behind weak flavor instabilities~\cite{Fiorillo:2024bzm, Fiorillo:2024uki, Fiorillo:2025npi}.

\section{Discussion and Summary}

\label{sec:summary}

Pendulum-like behavior turns out to be a generic feature of ensembles of interacting spins. A variety of interaction structures, among themselves and sometimes additionally with an external field, have appeared in the literature on collective neutrino flavor evolution, with each case often seeming like a new surprise.

We have reviewed the existing cases of the slow, fast, and single-wave systems and identified the common structure that makes pendulum-like behavior inevitable in minimal systems consisting of three vector variables. If one of these variables is conserved and another has a conserved length, the only possible interaction structures must be pendulum-like when viewed from the perspective of a suitably corotating observer. 

By contrast, if we extend these minimal systems to a larger ensemble of spins, or even to a continuum, the occurrence of bimodal self-maintained coherence is far less generic. Among the known examples, the slow homogeneous and isotropic, and fast homogeneous axially symmetric systems allow for such collective behavior, whereas the single-wave system does not.

In the context of collective neutrino flavor evolution, these mechanical spin systems arise only after imposing strong symmetry assumptions that reduce the phase space to a single parameter, which we denote by $v$. Moreover, other simplifying assumptions may be needed, such as the absence of refraction in a matter flow~\cite{Fiorillo:2023hlk}, which induces an additional vector in the EoMs, breaking the technical integrability of the system.

Whether such highly restricted systems are of practical relevance in realistic dense neutrino environments remains an open question. However, in the context of weak instabilities, we have recently argued~\cite{Fiorillo:2026vfo} that the single-wave system may serve as a useful proxy for the nonlinear saturation following an initial instability within a narrow range of wave numbers. 

Regardless of their practical applicability in neutrino physics, the flavor pendulum has long been a reference case of collective flavor conversion, and identifying its underlying abstract structures is therefore an interesting and worthwhile exercise.

\section*{Acknowledgments}

DFGF was supported by the Alexander von Humboldt Foundation (Germany) while most of this research was performed. GGR acknowledges partial support by the German Research Foundation (DFG) through the Collaborative Research Centre ``Neutrinos and Dark Matter in Astro- and Particle Physics (NDM),'' Grant SFB--1258--283604770, and under Germany’s Excellence Strategy through the Cluster of Excellence ORIGINS EXC--2094--390783311.

\bibliographystyle{JHEP}
\bibliography{References}

\end{document}